# Building Blocks of Physical-layer Network Coding

Jianghao He and Soung-Chang Liew, *Fellow*, *IEEE*

*Abstract*- This paper investigates the fundamental building blocks of physical-layer network coding (PNC). Most prior work on PNC focused on its application in a simple two-way-relay channel (TWRC) consisting of three nodes only. Studies of the application of PNC in general networks are relatively few. This paper is an attempt to fill this gap. We put forth two ideas: 1) A general network can be decomposed into small building blocks of PNC, referred to as the PNC atoms, for scheduling of PNC transmissions. 2) We identify nine PNC atoms, with TWRC being one of them. Three major results are as follows. First, using the decomposition framework, the throughput performance of PNC is shown to be significantly better than those of the traditional multi-hop scheme and the conventional network coding scheme. For example, under heavy traffic volume, PNC can achieve 100% throughput gain relative to the traditional multi-hop scheme. Second, PNC decomposition based on a variety of different PNC atoms can yield much better performance than PNC decomposition based on the TWRC atom alone. Third, three out of the nine atoms are most important to good performance. Specifically, the decomposition based on these three atoms is good enough most of the time, and it is not necessary to use the other six atoms.

*Index Terms*- Physical-layer Network Coding, Wireless Scheduling, Multi-hop Wireless Networks.

## I. Introduction

Since PNC was conceived [1], it has developed into a subfield of network coding under intensive research. The existing works on PNC can be grouped into three tracks depending on their orientation: 1) Communications; 2) Information Theory; and 3) Networking. Within these three fields of study, there are relatively few works under Networking, and most of the investigations in the other two tracks have focused on the simplest setup in which PNC can be applied, namely the two-way-relay channel (TWRC). Studies of the application of PNC in general networks are relatively few. To fill this gap, we raise two questions:

1. What are the fundamental building blocks of PNC?
2. How can these building blocks be put together to build large-scale general networks?

Naturally, the first question to ask is (i) whether there are other interesting PNC building blocks besides TWRC. A second question is (ii) how can these building blocks be used to decompose the transmission scheduling problem in wireless networks.

To set the context of our work, let us illustrate with an example. In a large general network, there could be many end-to-end traffic flows. Different flows may cross paths at various relay nodes. Fig. 1 shows five end-to-end flows in a network. From the local perspective of a relay node, each flow traversing it comes from a one-hop neighbor node and is destined for another one-hop neighbor. For example, for the relay labeled R in Fig. 1, there is a flow from neighbor A to neighbor C. The relay and its one-hop neighbors form a "local network". Two examples of

• Both authors are with the Department of Information Engineering, The Chinese University of Hong Kong, Hong Kong Special Administrative Region, China. E-mail: hjh010@ie.cuhk.edu.hk; soung@ie.cuhk.edu.hk

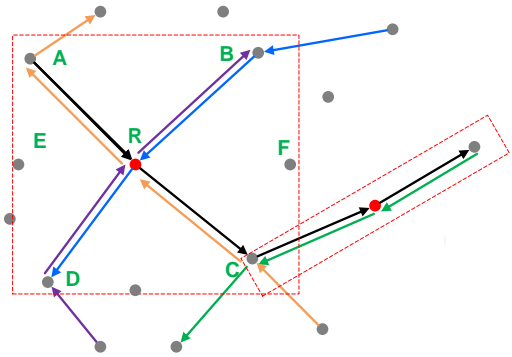

Fig. 1. A general network example with end-to-end flows.

| | TWRC | Hexagonal | Triangle | Others |
|---|---|---|---|---|
| SNC | 12 | 14 | 16 | X |
| PNC | 1,13 | 15 | X | X |

TABLE 1 STUDIES OF INDIVIDUAL BUILDING BLOCK

local networks are enclosed in the two rectangular frames in Fig. 1, where the relays are colored red. The sources and the destinations of traffic flows within a local network are one-hop neighbors of the associated relay.

In the context of the overall network, this paper assumes that the routes taken by the flows are predetermined by a separate routing algorithm. This paper focuses on how to schedule the transmissions of flows given their fixed routes. In a general network, the number of time slots needed to satisfy the traffic demands of all flows is often determined by the "bottleneck" relay with the most traffic crossing it. In this paper, we are interested in the scheduling problem faced by the bottleneck relay.

Returning to Fig. 1, consider the left local network enclosed in the square frame and suppose that node R is the bottleneck relay. There are four flows crossing the relay. Suppose that there are two packets to be delivered from node A to node C; two packets from C to A; one packet from B to D; and one packet from D to B. We note that there are two TWRCs embedded in this local network: A-R-C and B-R-D. As explained in [1], with the TWRC structure, PNC requires two time slots for two nodes to exchange one packet with each other. For example, in Fig. 1, two time slots are needed to deliver one packet from B to D and one packet from D to B under the TWRC construct. Thus, for the six packets above, using PNC TWRC decomposition, we need two time slots to deliver the two packets between B and D, and additional four time slots to deliver the four packets between A and C. Overall, TWRC decomposition requires a total of six time slots.

A solution requiring only five time slots is as follows. Suppose that A and C can overhear the transmissions of B and D, and B and D can overhear the transmissions of A and C. The lower three diagrams in Fig. 2(f) show a three-slot transmission pattern for A and C to exchange one packet with each other, and for B and D to exchange one packet with each other. For simplicity, this paper uses the regular font to label nodes and the italic font to label packets sent by the nodes. As shown, in the first time slot, nodes A and C transmit packets *A* and *C*, respectively. With the PNC mechanism [1], nodes R, B, and D receive packet (Section IV discusses the physical-layer operations needed for the decoding of $A \oplus C$). In the second time slot, nodes B and D transmit packets B and D, respectively. Nodes R, A, and C receive packet $B \oplus D$. In the third time slot, relay R XORs $A \oplus C$ and $B \oplus D$ to transmit $A \oplus B \oplus C \oplus D$, which is received by nodes A, B, C, and D. Each of the nodes A, B, C, and D could extract its target packet. For example, the target packet of node A is packet *C*. It can extract packet *C* by $(A \oplus B \oplus C \oplus D) \oplus (B \oplus D) \oplus A = C$. After the above three time slots, there is still one packet from A to C and one packet from C to A to deliver. We could use the TWRC A-R-C to deliver these remaining packets in two time slots. Thus, totally five time slots are consumed.

The bidirectional cross structure in Fig. 2(f) is a PNC building block that is different from TWRC. Note that the three-slot transmission pattern is distinct from the pair of two-slot transmission patterns of the TWRCs A-R-C and B-R-D, although they deliver the same four packets. We emphasize that the three-slot transmission pattern cannot be further broken down to an assembly of transmission patterns of other PNC building blocks. In this paper, we identify nine PNC building blocks (see Fig. 2) and refer to them as PNC atoms because the transmission pattern of each of them cannot be decomposed into those of other PNC atoms. The previous two paragraphs show that decomposition based on bidirectional cross and TWRC is more efficient than decomposition based on TWRC alone. A main result of this paper is that decomposition based on a variety of different PNC atoms is in general much more efficient than decomposition based on TWRC alone. Furthermore, our study indicates that with the decomposition framework, PNC scheduling can significantly outperform scheduling based on the traditional multi-hop scheme (non-NC) and the straightforward network coding scheme (SNC) that performs network coding at the higher layer. For example, under heavy traffic volume, compared with non-NC, PNC achieves roughly the same throughput gain (~100%) in general networks as it does in the simple TWRC network.

Our contributions can be summarized as follows (answering Questions 1 and 2 respectively):

1. TWRC is only one of many PNC atoms. Besides PNC TWRC, we identify eight other PNC atoms. We present formal definitions for the nine PNC atoms and formulate the PNC scheduling problem as a linear program based on the atoms.

2. A general network can be decomposed into small building blocks of PNC for scheduling purpose. We show that their joint use can boost throughput in wireless networks significantly.

The rest of this paper is organized as follows. Section II describes related work. Section III defines a PNC atom formally as an entity consisting of three attributes. Section IV overviews physical-layer (PHY) issues related to PNC atoms. Section V shows that the problem of PNC scheduling can be formulated as a linear program (LP) under our decomposition framework. The LP minimizes the number of time slots needed to satisfy the traffic demands of flows traversing the bottleneck relay. Section VI investigates various variants of our decomposition scheme and the relative importance of different atoms. Section VII proposes a MAC protocol for the coordination of PNC transmissions scheduled according to the decomposition result. Section VIII concludes this paper.

## II. RELATED WORK

Although this paper focuses on MAC issues related to PNC atoms, many of our assumptions are based on the physical-layer operations of PNC systems. Many PHY challenges for PNC have been tackled by prior work. Overview of PHY-layer issues can be found in the tutorial paper [3]. References [4-8] address synchronization issues; [3] and [9] consider general PNC mapping issue; and [10] and [11] address channel estimation. Section IV gives a brief discussion on the PHY-layer operations upon which the MAC investigations of this paper are based.

We researched earlier studies of network coding in wireless networks and found that the concept of *building blocks* has not been propounded previously. Here, we extrapolate the potential relevance of the earlier studies to our new concept of building blocks, where applicable.

Our findings are summarized in Table 1. Entries marked with "X" are areas that have not been studied previously. Table 1 lists prior works related to the studies of individual building blocks. Ref. [12] considered the use of SNC (i.e., non-physical-layer network coding) in TWRC. Ref. [1] proposed PNC for TWRC. We omit most of the subsequent studies on PNC TWRC in Table 1 due to their sheer volume. The study of PNC TWRC culminated in [13], in which it was shown that with the use of lattice code, information rates within 1/2 bit of the cutset upper bound of TWRC information capacities can be achieved with PNC. This firmly established the fundamental value of PNC in TWRC. Ref. [14] considered a hexagonal structure in which six nodes communicate with the assistance of a relay using SNC. Ref. [15] considered the same hexagonal building block for PNC. It is assumed that each node can simultaneously receive from up to three other nodes and can perform PNC decoding based on the received composite signal. In this paper, for practical reasons, we limit ourselves to the case where a node can simultaneously receive from at most two other nodes. Ref. [16] described a general structure which includes the triangular structure in which three nodes communicate through a relay; however, SNC rather than PNC was considered. The important point to draw from Table 1 is that there have been very few studies of PNC building blocks beyond TWRC.

There has also been work considering PNC broadcast, in which the same information from each node is to be relayed to all other nodes via a relay (see, for example, [17]). Furthermore, Ref. [25] proposed a PNC multicast protocol considering multiple relays. In the current paper, however, we are interested in PNC unicast with one relay.

A simple version of PNC called Analog Network Coding (ANC) that employs amplify-and-forward strategy (rather than XOR-and-forward as in PNC) was proposed in [18]. In addition to TWRC, a cross structure was considered in [18]. In [24], a MAC layer algorithm for ANC was presented.

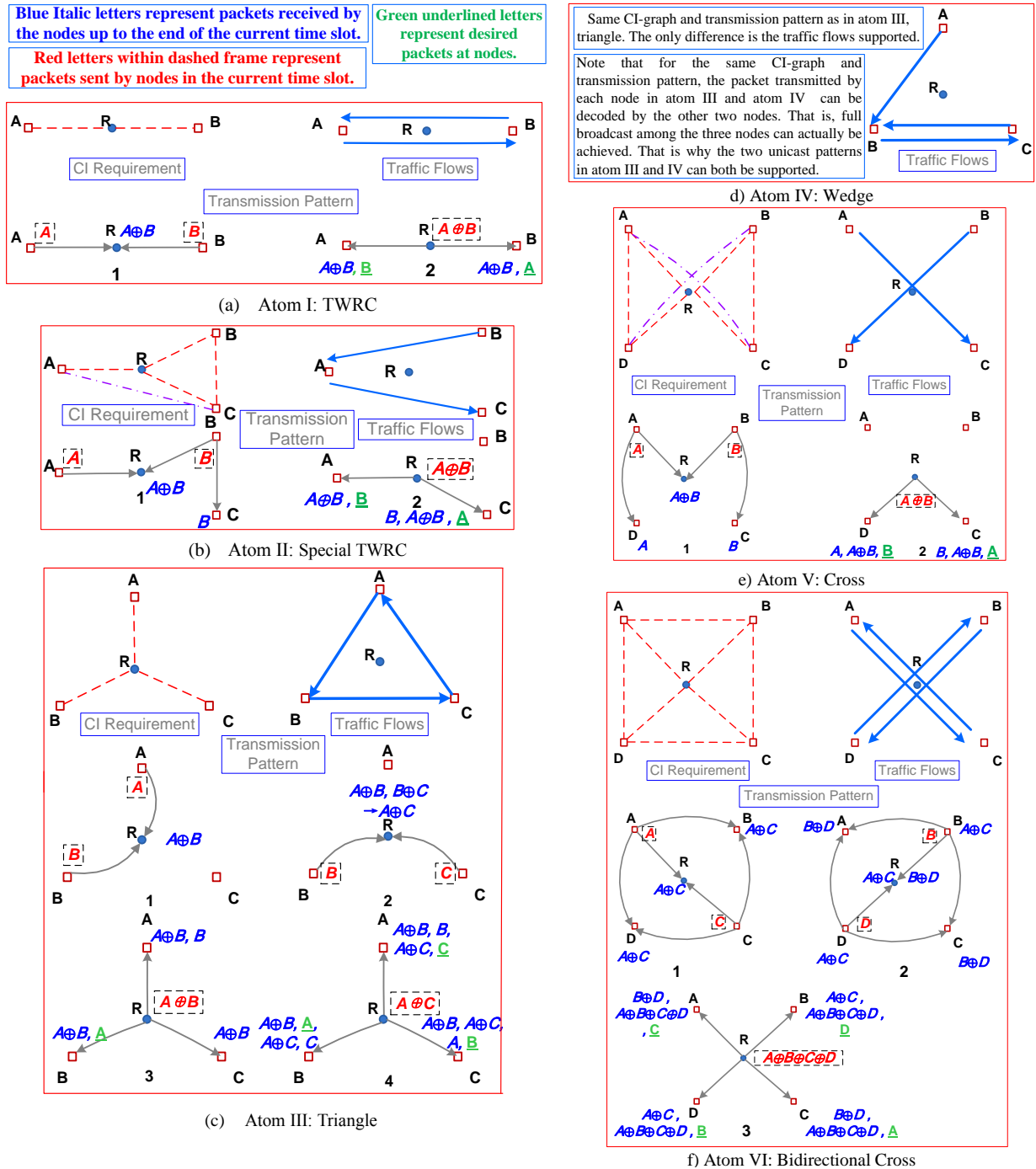

(a) Atom I: TWRC
(b) Atom II: Special TWRC
(c) Atom III: Triangle
(d) Atom IV: Wedge
(e) Atom V: Cross
(f) Atom VI: Bidirectional Cross

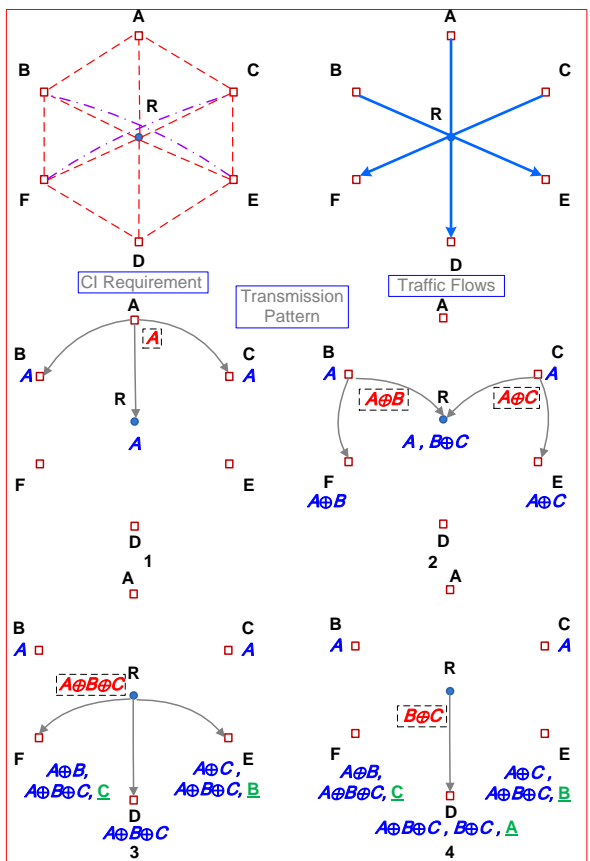

g) Atom VII: Star I (Asymmetric Star)

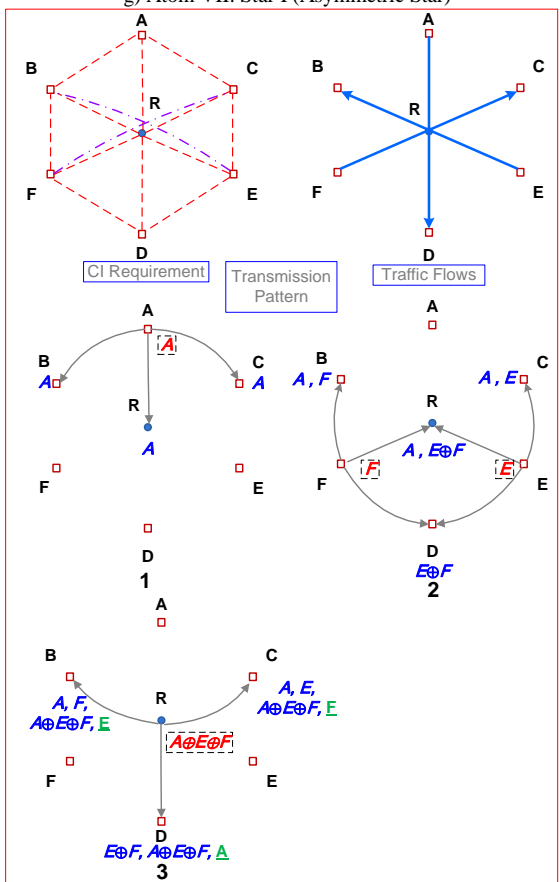

h) Atom VIII: Star II (Symmetric Star)

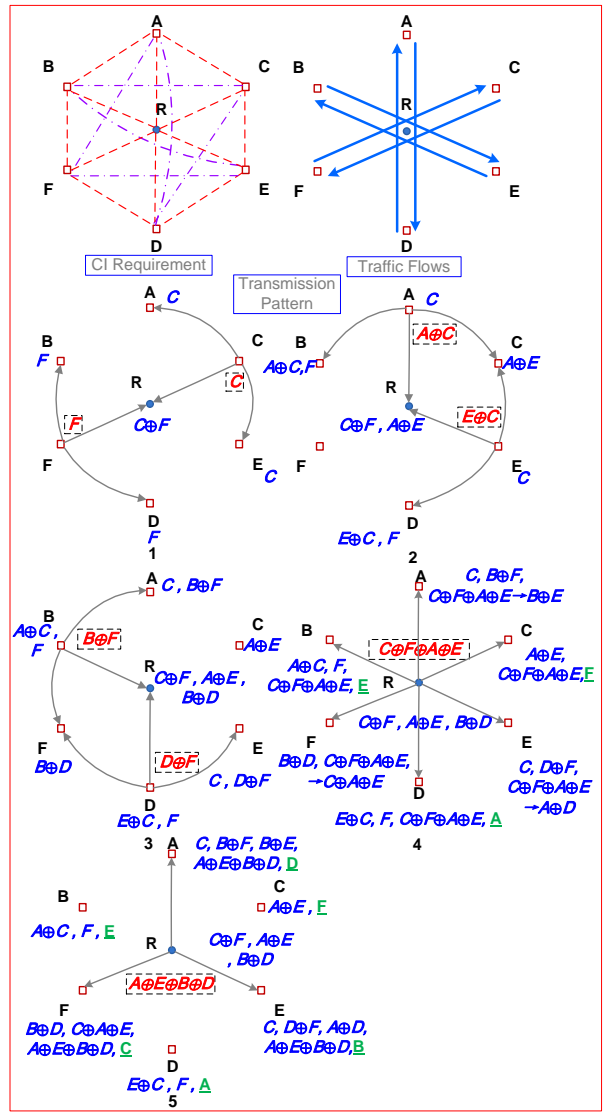

(i) Atom IX Star III (Bidirectional Star)

Fig. 2. Nine different atomic PNC building blocks.

## III. PNC ATOMS

This section defines PNC atoms. Broadly speaking, a PNC atom specifies how the traffic among a group of nodes (referred to as the *peripheral nodes*) around a relay can be delivered via the relay. Given the local topology around a relay, we can form a number of PNC atoms to facilitate the traffic delivery.

### A. Attributes of PNC Atoms

Each PNC atom is defined by three attributes. Each attribute can be described by a diagram, as shown in Fig. 2. We detail the three attributes below:

*1) Connectivity-Interference (CI) Requirement*

The first attribute of an atom is the connectivity and interference relationships among its peripheral nodes. These relationships can be modeled by a graph, referred to as the *CI requirement graph* (CI-graph). These requirements must be satisfied for an associated transmission pattern (described in *(3)*) to work.

*Connectivity Requirement*: Two nodes are connected by a connectivity edge (C-edge), represented by a red dashed line in Fig. 2, if they are required to be within the transmission range of each other. We make the following assumption:

**Assumption 1**. *The relay R in a PNC atom is within the transmission range of all peripheral nodes.*

This is a reasonable assumption: if a relay were not connected to the peripheral nodes, it could not function as their relay. Two peripheral nodes, on the other hand, may or may not be connected; a C-edge between two peripheral nodes indicates that they could overhear each other.

*Interference-free Requirement*: Two nodes are connected by an interference-free edge (I-edge), represented by a purple dashed-dotted line in Fig. 2, if they must be out of the interference range of each other. This requirement will guarantee that one node can successfully receive a packet without being interfered by the transmission of the other node. For example, in Fig. 2(e), nodes A and B simultaneously transmit in the first time slot; D must be out of the interference range of B for it to receive successfully from A. In Fig. 2, atoms I, III, IV and VI do not have I-edges in their CI-graphs, while atoms II, V, VII, VIII and IX do.

Note that the CI-graph only states the required conditions for the associated transmission pattern (described in *(3)*) to work. The absence of a C-edge between two nodes does not necessarily mean that they are out of the transmission range of each other; it just means it is not known whether two nodes can hear each other. Similarly, the absence of an I-edge between them does not necessarily means that they are within the interference range of each other. The absence of an edge between two nodes only means that the associated relationship does not matter as far as the transmission pattern is concerned.

*2) Traffic Flows*

The second attribute of an atom is the traffic that it can deliver. In an atom, the relay assists the delivery of the traffic among the peripheral nodes. The traffic flow from a source peripheral node to a destination peripheral node that the atom can support is represented by a blue directed line in the traffic flow diagram in Fig. 2. We make the following assumption about the traffic flows:

**Assumption 2.** *For each flow in a PNC atom, the destination is outside the transmission range of the source.*

This is a reasonable assumption of the physical situation of interest to us: if the destination were within the transmission range of the source, the source could transmit directly to it without the help of the relay.

*3) Transmission Pattern*

The third attribute of an atom is the transmission pattern (and the accompanying PNC mechanisms) used to deliver one packet for each traffic flow of the atom. As shown in Fig.2, a transmission pattern consists of a number of time slots. In each time slot, some nodes transmit and some nodes receive. The packets transmitted by the transmitting nodes and the packets received by the receiving nodes are specified in the transmission pattern. To simplify the reception (and overhearing) mechanism at the nodes, we make the following assumption:

**Assumption 3.** *Half duplexity: a node cannot transmit and receive at the same time.*

The relay node participates either as a receiving node or a transmitting node in all time slots. Accordingly, we divide the transmission pattern into two phases: i) In the uplink phase, the relay receives. A source node of a flow sends either its native packet or a network-coded packet consisting of its native packet mixed with overheard packets from other source nodes. ii) In the downlink phase, the relay transmits network-coded packets.

The existence of a C-edge between two nodes means that they are neighbors who can hear each other. We assume that if two neighbors of a node transmit in the same time slot, provided the node is not transmitting itself, it can derive a network-coded packet from simultaneously received signals using the PNC mechanism [1], [3].

A transmitting node may combine several previously received packets to transmit a network-coded packet. For example, in the last time slot in Fig. 2(f), relay R combines $A \oplus C$ with $B \oplus D$ to transmit $A \oplus B \oplus C \oplus D$.

In practice, it is complex for the relay to decode a network-coded packet of three or more simultaneous signals, although this is possible in theory. Hence, we make the following assumption:

| TABLE 2 | TIME SLOTS CONSUMED BY PNC, SNC AND NON-NC SCHEMES UNDER EACH ATOM | | | | | | | | |
|---------|---|---|---|---|---|---|---|---|---|
| Scheme | Atom Type | | | | | | | | |
|  | I | II | III | IV | V | VI | VII | VIII | IX |
| *PNC* (TS) | 2 | 2 | 4 | 4 | 2 | 3 | 4 | 3 | 5 |
| *SNC* (TS) | 3 | 3 | 5 | 5 | 3 | 5 | 5 | 4 | 8 |
| *Non-NC* (TS) | 4 | 4 | 6 | 6 | 4 | 8 | 6 | 6 | 12 |

*TS*: number of time slotsots.

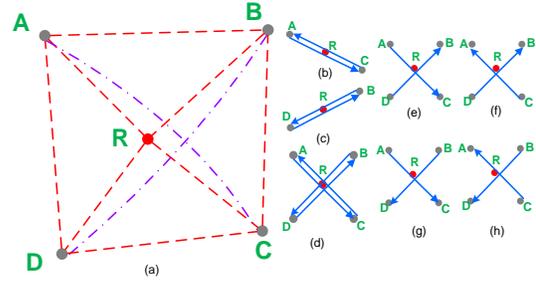

Fig. 3. The CI-graph of an example local network.

**Assumption 4.** *In the uplink phase, at most two source nodes transmit together in the same time slot.*

In general, the transmission pattern of an atom may be such that a network-coded packet is to be decoded by more than one node in a time slot. For example, in the first time slot in Fig. 2(f), nodes R, B and D must decode packet $A \oplus C$. In Section IV, we will provide further details to readers who are not familiar with the PHY-layer operations needed for the decoding of network-coded packets in PNC.

*B. Nine PNC Atoms*

Fig. 2(a) to (i) present nine PNC atoms together with their three attributes. The nine atoms are sorted by the number of the peripheral nodes, from small (two) to large (six). We can divide these nine atoms into four types according to their structures. In the following, we explain the operation of one atom for each type in detail (the operations of the other atoms can be extrapolated from the explanation below):

1) *TWRC:* includes Atoms I and II. TWRC is the well-known structure to which PNC was applied when the concept of PNC was first proposed [1]. We skip the explanation here. The reader is referred to Section I, where the illustration example in the eighth paragraph includes the explanation of the operation of TWRC (the 8[th] paragraph).

2) *Triangle:* includes Atoms III and IV. Here, we explain how atom III works. As shown in Fig. 2(c), there are three peripheral nodes A, B and C around the relay. They could not directly communicate with other. There are three traffic flows: from A to B, from B to C and from C to A. With the non-NC scheme, six timeslots are needed in total: three to send packets from the sources to the relay, and three to forward packets from the relay to the destinations. With PNC, in the first timeslot, nodes A and B transmit packets *A* and *B* respectively. Node R receives packet $A \oplus B$. In the second timeslot, nodes B and C transmit packets *B* and *C* respectively. Relay R receives packet $B \oplus C$, then extracts packet $A \oplus C$ by $(A \oplus B) \oplus (B \oplus C) = A \oplus C$. In the third timeslot, relay R broadcasts $A \oplus B$ to nodes A, B and C. Node B then extracts its target packet *A* by $B \oplus (A \oplus B) = A$. Node C stores the received $A \oplus B$. In the fourth timeslot, relay R broadcasts $A \oplus C$ to nodes A, B and C. Node A extracts its target packet $C$ by $A \oplus (A \oplus C) = C$. Node C then extracts its target packet $B$ by $(A \oplus B) \oplus (A \oplus C) \oplus C = B$. PNC requires only four timeslots.

3) *Cross:* includes Atoms V and VI. The reader is referred to the ninth paragraph of Section I on how atom VI operates.

4) *Star:* includes Atoms VII, VIII and IX. Here, we explain how atom VIII works. As shown in Fig. 2(h), there are six peripheral nodes and three traffic flows between them: one from nodes A to D, one from E to B and one from F to C. Since each pair of source and destination cannot communicate with each other directly, with the non-NC scheme, a total of six timeslots are needed. With PNC, in the first timeslot, node A transmits packet $A$, which is received by nodes R, B and C. In the second timeslot, nodes E and F transmit packets $E$ and $F$ respectively. Both nodes R and D receive packet $E \oplus F$. Since node B is within the transmission range of node F but out of the transmission range of node E, node B overhears packet $F$. Similarly, node C can overhear packet $E$. In the third timeslot, relay R XORs $A$ and $E \oplus F$ and transmit $A \oplus E \oplus F$, which is received by nodes B, C and D. Each of the nodes B, C and D could extract its target packet. For example, node B could extract its target packet $E$ by $A \oplus (A \oplus E \oplus F) \oplus F = E$. PNC requires only three timeslots.

Table 2 lists the numbers of time slots needed by the PNC atoms. Also shown are the numbers of time slots needed for the non-NC and the SNC schemes. For SNC, we study the nine atoms modified for SNC. There are two main differences between SNC and PNC atoms: 1) SNC atom consumes more time slots, e.g., SNC TWRC requires three time slots; 2) there is no interference-free requirement in the SNC atoms. We see from Table 2 that the ratios of the numbers of time slots of the PNC atoms and Non-NC atoms range from 4:6 to 3:8, corresponding to throughput gains ranging from 50% to 167%.

*C. Atom Class, Atom Instance, and Decomposition*

In the lingo of object-oriented programming, the nine atoms depicted in Fig. 2 are nine "class" specifications. Each atom instance fits into the template of an atom class.

In a local network, there could be several atom instances belonging to different classes. For example, Fig. 3(a) shows the structure of a local network with four peripheral nodes around a relay. The red dashed lines represent the connectivity requirements and the purple dashed-dotted lines represent the interference-free requirements between nodes. Embedded in this topology are seven atom instances. There are two instances of atom I: their traffic flow diagrams are shown in Fig. 3(b) and (c). There is one instance of atom VI: its traffic flow diagram is shown in Fig. 3(d). There are four instances of atom V: their traffic flow diagrams are shown in Fig. 3(e) to (h).

Suppose that the traffic to be delivered across the network in Fig. 3(a) is as follows: 3 traffic units (packets) from A to C; 4 from C to A; 2 from B to D; and 1 from D to B. Accordingly, let us denote the numbers of packets to be delivered by the vector (3, 4, 2, 1). We could use atom instance (d) once, atom instance (h) once, and atom instance (b) twice to deliver the packets: after using atom instance (d) once, the remaining packets are (2, 3, 1, 0); after using atom instance (h) once, the remaining packets are (2, 2, 0, 0); after using atom instance (b) twice, the remaining packets are (0, 0, 0, 0). The total time slots used is $3 + 2 + 2*2 = 9$. As long as the ratios of the traffic between the nodes are 3: 4: 2: 1, we could use the above schedule repeatedly to deliver the traffic. We refer to scheduling traffic this way using atom instances as "decomposition". In general, different decompositions are possible to meet the same traffic demands. The more efficient decompositions use fewer time slots. Section V shows how to formulate the decomposition optimization problem as a linear program (LP).

Let us revisit the atom classes to make a point regarding isomorphism between classes. Recall that each of the atoms in Fig. 2 is specified by a configuration consisting of a CI-graph, a traffic flow diagram, and a transmission pattern. There could be configurations that are isomorphic to each other. Two configurations are isomorphic if by permuting the labels of the nodes, we can transform one configuration to the other (i.e., the CI-graph, traffic flow diagram, and transmission pattern are transformed to those of the other configuration). Isomorphic configurations are not distinct atom classes. The atom classes in Fig. 2 are not isomorphic. Fig. 3(e) to (h) are atom instances belonging to the same atom class V: by relabeling nodes, we could fit them into the template of atom class V in Fig. 2(e).

For each atom class in Fig. 2, the transmission pattern shown is not the only possible pattern. However, the transmission patterns for all the atom classes given in Fig. 2 are optimal in the sense that there are no other patterns consuming fewer time slots. The proofs are given in Appendix A.

## IV. PHY ISSUES

To achieve the gain of PNC presented in Section III, a PNC atom needs to successfully decode the desired packets at the PHY layer as required by its transmission pattern.

This section overviews the related PHY issues, pointing the readers to key references for further details. This section does not contain new contributions; rather, its goal is to provide the context and the PHY-layer justifications for our MAC-layer and network-layer operations here. Readers who are already familiar with the various PHY issues of PNC can skip this section and proceed directly to Section V without losing.

Most prior work on PHY layer of PNC focuses on TWRC (i.e., atom I). To date, most PHY challenges for PNC TWRC have been tackled. The other eight PNC atoms do not cause many new difficulties beyond those already addressed. For each following subtopic, we will first overview prior PHY results on TWRC; then, we will explain how the results or insights of the TWRC studies can be directly applied or extended for application in the other atoms covered by our paper here. Generally speaking, the main PHY issues of PNC TWRC are related to how the relay maps the overlapping signals to a network-coded packet (referred to as the *PNC mapping*) [3] in the uplink phase. We note the following two points with respect to the other eight atoms:

Let us consider the other PNC atoms. We note the following two points:

1) In the uplink phase, some non-TWRC atoms may require more than two peripheral nodes need to send information to the relay (e.g., nodes A, B, C, and D in atom VI). For simplicity of PNC decoding, we introduced assumption 4 in Section III.A.3 to restrict the number of simultaneously transmitting nodes to no more than two. If more than two peripheral nodes need to transmit to the relay, we use more than one time slot for the uplink phase (e.g., in atom VI, nodes A and C transmit in one time slot, and nodes B and D transmits in another time slot). Consequently, PNC mappings at the relay of the other atoms are similar to that of TWRC.

A non-TWRC atom, however may require nodes other than the relay to also decode the network-coded packet through the overhearing process (e.g., in atom VI, in the first time slot of the transmission pattern, besides relay R, nodes B and D also need to decode packet $A \oplus C$). However, PNC mappings at the overhearing nodes are similar to that of TWRC.

2) In the downlink phase, although we can allow a peripheral node to transmit together with the relay, doing so cannot save time slots (a proof is given in Appendix A); on the other hand, more complicated interference-free requirements are will be required for proper decoding at the receiver. Hence, for simple operation, only the relay transmits in our atoms. Furthermore, the relay only transmits one packet (network-coded packet) in a transmission. As far as the transmission from the relay to each and every of the target receiving node is concerned, we have a conventional point-to-point transmission. Thus, for the remainder of this section, we only overview the PHY issues related to the decoding of network-coded packets at a receiver during the uplink phase, with illustration based on atom VI when necessary.

*A. Synchronization*

In the uplink phase, the received signal at relay R from nodes *i* and *j* can be expressed as

$$y_R = h_{iR} x_i + h_{jR} x_j + n_R \tag{1}$$

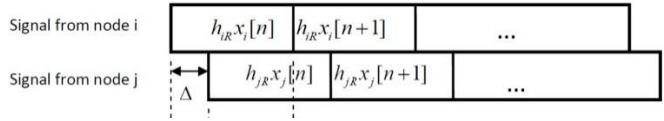

Fig. 4. Symbol offset between the signals from nodes *i* and *j* in TWRC.

where $h_{iR}$ is the complex number denoting the channel gain (fading coefficient) from node *i* to relay R, and $n_R$ is the Gaussian noise.

Two major asynchronies at a PNC receiver are phase asynchronies and symbol asynchronies [3] due to the fading. Phase asynchrony can be captured into (1) with $\angle h_{iR}/h_{jR} \neq 0$. Symbol asynchrony refers to the symbols of different transmitters being not aligned at the receiver due to different arrival times. The situation is depicted in Fig. 4, where the baseband signals form nodes *i* and *j* are shown.

Fortunately, the penalties due to phase and symbol asynchronies can be nullified to a large extent with the incorporation of channel coding into the PNC system. Lu and Liew [4] investigated a general framework for decoding at the receiver based on belief propagation (BP) when $0 < \Delta < 1$. For channel-coded PNC, with their BP method, both symbol and phase asynchronies actually improve the system performance compared with the perfectly synchronous case. Yang and Liew [7] considered a three-layer decoding framework for asynchronous convolutional-coded PNC systems when $\Delta \geq 1$. The performance degradation due to asynchronies is within 1dB.

Additionally, [5] reported the first PNC prototype based on OFDM, which obviates the need for tight synchronization. With OFDM, if the relative sample delay offset of the simultaneously transmitting nodes is within the length of the Cyclic Prefix (CP), then in the frequency domain, the symbols will experience no offset. To ensure within-CP arrivals, [6] made use of beacons as in the 802.11 standard to coordinate the local timers of the transmitters.

For OFDM system, there is also the issue of carrier frequency offsets (CFO), which can cause inter-carrier interference (ICI). However, CFO asynchrony does not pose major problems in the PNC prototype in [6] either. The work [6] made use of CFO precoding to reduce the relative CFO to an extent that the residual CFO causes only insignificant ICI.

Consider atom VI. For the first time slot, it will be difficult to achieve synchronized receptions at all of nodes R, B and D (Fig. 2(f)). However, when channel coding and/or OFDM is used, asynchronous decoding at the nodes are possible with the methods expounded in [5]-[7].

We also assume the transmit power has been factored into the $h_{iR}$. For near-far problem, we can use power control to guarantee that the packets from different end nodes can arrive at relay R with the same amplitude. The other solution is to use multiuser detection technique (e.g., successive interference cancellation) for directly detecting the two packets (not the XORed form) before XORing them together [3]. Yet another solution is to use perform PNC mappings that can deal with unequal powers from the two users [26]. Our paper here does not go into the details of PHY-layer signal processing, focusing instead on what can be done at the MAC layer given than the PNC mechanism can be realized by the PHY layer.

## B. General PNC Mapping

PNC mapping maps $y_R$ in (1) to a target symbol $z_R$ (e.g., $z_R = x_i \oplus x_j$) for broadcast back to nodes $i$ and $j$. In general, PNC mapping is not limited to just the XOR mapping. In the broadest scope of PNC [3], PNC mapping is a network coding function: $z_R = f(x_i, x_j)$. Nazer and Gastpar [9] used lattice codes in a compute-and-forward framework so that the relay can decode a linear function of transmitted *codewords* according to the channel coefficients. Roughly speaking, the relay attempts to map the received signal to $z_R = \alpha_{iR} x_i + \alpha_{jR} x_j$ where $\alpha \in \mathbb{Z}$ and pass this *codeword equation* to the end nodes for final decoding. Such more general linear mappings will allow for better performance in the high SNR regime.

In this paper, to simplify discussion, we assume $z_R = f(x_i, x_j) = x_i \oplus x_j$. More generally, we do not restrict ourselves to bit-wise XOR only. In other words, by the notation $x_i \oplus x_j$, we mean the general $z_R = \alpha_{iR} x_i + \alpha_{jR} x_j$.

## C. Channel Estimation

PNC mapping is predicated on the availability of accurate channel estimates. Channel estimation in PNC is particularly challenging because of the overlapping of signals from multiple users. To tackle this challenge, [10] established a joint channel estimation and channel decoding framework that combines the use of the expectation maximization (EM) algorithm and the belief propagation (BP) algorithm. The iterative algorithm enables accurate channel estimation and channel decoding for PNC mapping.

Consider atom VI. A subtlety that does not exist in TWRC is that for atom VI, during the uplink phase, several receiving nodes will experience different channel gains. Thus, the channel gains need to be estimated at the respective receivers. In atom VI, nodes R, B and D can just independently do the joint channel estimation and channel decoding for PNC.

## D. Error Control in PNC Systems

Error control is a fundamental functionality in communications systems with noise. Error control can be realized in two different ways:

*1) Forward error correction (channel coding):* For our PNC atoms, an underlying assumption is that link-by-link channel coding [3] is adopted. The goal is to channel-decode the signals not into individual messages of the two transmitters, but into a network-coded message [3]. Such link-by-link channel-coded PNC allows the relay to denoise the signals before forwarding the network-coded message along. Section 3.2 of [3] discusses several link-by-link PNC schemes.

*2) Automatic repeat request:* Despite forward error control, packets may still get corrupted and not receive properly. In Section VII.A, we will introduce an end-to-end acknowledgement (ACK) mechanism in MAC layer to ensure overall reliability of the PNC systems.

## V. SCHEDULING USING ATOM DECOMPOSITION

This section formulates the scheduling problem at a local network. In this and the next sections, we make the following assumption about the traffic demands at a local network:

**Assumption 5.** *In a local network, as far as the traffic routed through the relay R is concerned, there are no traffic demands between two peripheral nodes that are within the transmission range of each other.*

If there is traffic from one peripheral node to another peripheral node within its transmission range, it could transmit directly to that peripheral node without going through the relay. Our scheduling problem here focuses on the traffic routed through the relay R.

We tackle the scheduling problem in two steps: (i) identifying the atom instances in the neighborhood of relay R; (ii) formulating an LP (linear program) and solving it to find the amounts of airtime to allocate to the atom instances. In the following two sections, we detail these two steps.

### A. Identification of Atom Instances

Before we can go about identifying the atom instances, we need to first list the traffic flows between peripheral nodes that could be delivered via R (abbreviated as flows henceforth). Consider a relay R. Let $N_R$ be the number of R's neighbors. There are then $N_R(N_R-1)$ *possible flows* that may cross R. Let $F$ be the number of *potential flows* that will actually make use of R to communicate. In general, $F \leq N_R(N_R-1)$. The inequality is due to the fact that the distances between some pairs of nodes may be within the transmission range. The nodes of such a pair can communicate directly with each other, and we assume the associated flow will not use R as a relay for efficiency's sake (Assumption 5). Hence we denote a potential flow from node A to node B by A→B if they are outside the transmission range of each other. Let $F_R$ ($|F_R|=F$) be the set of potential flows. For example, for a local network with three peripheral nodes A, B and C and they cannot communicate directly with each other, we write $F_R$= {A→B, B→A, B→C,C→B, A→C,C→A}. Note that potential flows are ordered, so that A→B and B→A denote two distinct flows.

By "identifying atom instances", we mean finding the atom instances in the local network that fit into the templates of the nine atom classes. With respect to the traffic flow diagram of atom class $i$, let $M_i$ denote the number of flows in it (e.g., $M_{II}$=4). We could represent each atom instance by a set of node pairs, each corresponding to a flow. For example, we could represent a particular instance of atom-I (TWRC) class indexed as $j$, say $a_1^j$, by $a_1^j = \{A \rightarrow B, B \rightarrow A\}$.[1] Let $A_i = \{a_i^j \; \forall j\}$ be the set of all atom instances of class $i$ in the neighborhood of R. Furthermore, let us index the peripheral nodes by $k$=1 to $k=N_R$, and let $n_k$ be the $k_{th}$ node. We can then represent $A_i$ in the following form:

$$A_i = \{a_i^1, a_i^2, \ldots\} = \{\{n_1 \rightarrow n_3, n_2 \rightarrow n_5, \cdots, n_u \rightarrow n_v\}, \{n_1 \rightarrow n_3, n_2 \rightarrow n_5, \cdots, n_j \rightarrow n_k\}, \ldots\}$$

where there are $M_i$ flows in each inner brace (e.g., $\{n_1 \rightarrow n_3, n_2 \rightarrow n_5, \cdots, n_u \rightarrow n_v\}$).

For the overall atom-instance identification algorithm, the input is the flow (node-pair) set $F_R$, and the outputs are nine sets of atom instances, $A_1$ to $A_9$. The identification process is to find the atom instances that belong to each of the nine atom classes based on the given $F_R$.

To find the instances of $A_i$, for each combination of $M_i$ node pairs out of $C_F^{M_i}$ possible combinations[2] in the local network, we check whether the $M_i$ node pairs satisfy the CI requirement of PNC atom $i$. For illustration, in the following we elaborate the CI requirement check for atom class V in Fig. 2(e). Here, $M_V$=2. Consider two local flows in $F_R$ chosen for checking: (A, C) and (B, D). We want to see if these two flows constitute an instance of atom V by performing the following CI requirement check:

The CI requirement check is divided into connectivity check and interference-free check. For connectivity check, since relay R and its neighbor nodes are connected by definition (Assumption 1), we only need to check the connectivity between the peripheral nodes. With reference to the CI-graph of Fig. 2(e), note that nodes A and D are required to be connected, and nodes B and C are required to be connected. To check this, we look at $d_{AD}$ (distance between A and D) and $d_{BC}$ (distance between B and C). If both $d_{AD}$ and $d_{BC}$ are within the transmission range, they are connected. After the connectivity check, we move on to check the interference-free requirement.

The interference-free requirement check is performed with reference to the CI-graph of Fig. 2(e) as well. When nodes A and B simultaneously transmit to nodes D and C respectively, we require that B's transmission will not interfere with D's "overheard" reception from A, and A's transmission will not interfere with C's "overheard" reception from B. A typical method to determine the interference range of a link is to set an upper-bound distance proportional to the link length (the proportion is typically larger than one). For example, we compute $d_{BD}$ (distance between B and D) and $d_{AC}$ (distance between A and C). If both $d_{BD}$ and $d_{AC}$ are larger than $\alpha * d_{AD}$ and $\alpha * d_{BC}$, respectively, for some $\alpha > 1$, then the interference-free requirements are satisfied.

---

[1] Note that typically a flow (node pair) may belong to multiple atom instances (see Fig. 3 for an example).
[2] This means that the computational complexity of identifying atom instances is $O(F^{M_i}) = O(N_R^{2M_i}) = O(N_R^{12})$, where $\max_i M_i = M_{VIII} = 6$. This implies that this is not an NP-hard problem.

If the overall CI-requirement is satisfied, we then add the atom instance consisting of the node pairs (A, C) and (B, D) to $A_5$: $A_5 = \{(n_1 \rightarrow n_2, n_3 \rightarrow n_4), \ldots, (A \rightarrow C, B \rightarrow D)\}$.

In a similar way, we can obtain the atom instances of the other eight PNC atom classes. Based on all the atom instances identified, we then construct a PNC *atom database*. We represent the database by an $F \times \sum_{i=1}^{9} |A_i|$ incidence matrix **D**, in which each column corresponds to an atom instance, each row corresponds to a flow. In **D**, element $(i,j)=1$ if atom instance $j$ serves flow $i$, and element $(i,j)=0$ otherwise.

## B. Formulation of LP

Given **D** and the traffic demands between flows, the following LP finds the optimal schedule that consumes the least number of time slots to meet the traffic demands:

$$\min_{\mathbf{y}} \mathbf{b}^T \mathbf{y} \qquad \text{s.t.} \quad \mathbf{D}\mathbf{y} \geq \mathbf{c}, \mathbf{y} \geq 0 \qquad (2)$$

where **c** is an $F \times 1$ vector whose elements are integers representing the traffic demands of the $F$ flows crossing R; **y** is an $\sum_{i=1}^{9} |A_i| \times 1$ vector describing the airtime allocated to each of the $\sum_{i=1}^{9} |A_i| \times 1$ atom instances (i.e., the relative frequency the transmission pattern of an atom instance is executed); **b** is an $\sum_{i=1}^{9} |A_i| \times 1$ vector describing the numbers of time slots required by the transmission patterns of the atom instances; **D** is an $F \times \sum_{i=1}^{9} |A_i|$ matrix containing the atoms identified through the process described in Part A. In particular, **D** is fixed here and not variables to be optimized in (2).

**Remark**: Strictly speaking, the problem we have is an integer linear program (ILP) rather than LP, because **y** corresponds to the numbers of times atom instances are scheduled. ILP is a much tougher problem than LP. Fortunately, our simulation results indicate that the solutions of LP (2) are integral most of the time. The intuition of this is as follows: later in Section VI.D, we show that atoms I, II and V are the most important atom classes responsible for good performance. In particular, these three atoms are most prevalent (easy to be identified) and their combination can replace the remaining atoms with little throughput degradation. Hence, they are scheduled most often in the optimal LP solutions. Furthermore, if **D** only contains atom instances that have two traffic flows such as atoms I, II and V (i.e., we exclude atoms with more than two traffic flows in the decomposition), then it can be proved that **D** is totally *unimodular.* When **D** is totally *unimodular*, there is an integral optimal solution to LP (2). For example, simplex method will automatically an integral optimal solution. In this work, when we have a non-integral solution, we simply convert it to an integral (possibly suboptimal) solution by rounding. Doing this can significantly reduce the computational complexity. In particular, the complexity of the rounding algorithm is just $O(n)$ where $n$ is the total number of atom instances. If we use simplex method to solve LP, *on average*, the complexity is $O(n^3)$ (previous studies show that in the worst case, simplex method is of exponential complexity in the worst case, but the worst case occurs rarely in practice). So the total *average* cost is just $O(n^3)$, while the complexity of directly solving ILP by exhaustive search is $O(m^n)$, where $m$ is the largest traffic demand among all traffic flows.

In summary, the overall schedule procedure is as follows: Given a local network and traffic demands **c** between nodes, we first identify the atom instances within the local network to form an identified atom database **D**. Then we formulate LP (2) based on the traffic demands **c** and the identified atom database **D**. Solving LP (2), gives us the optimal airtime allocation **y** that indicates which atom instances should be scheduled and how many times they should be scheduled. An order of who should transmit at what time can then be constructed. Following this schedule, all traffic can be delivered in minimal amount of time.

## VI. PERFORMANCE EVALUATION

This section is devoted to the performance study of the decomposition method described above. Extensive simulation experiments have been conducted for the evaluation of the relative importance of different atoms.

### A. Definition of Traffic Volume

Let $K$ be the number of traffic units passing through the relay between the peripheral nodes. We refer to $K$ as the *traffic volume*. Our simulations are conducted with different $K$ values. Under packet switching, we can interpret $K$ as the number of packets being scheduled in a round. If $T_i$ time slots are needed for round $i$ to deliver the $K$ packets, then the throughput averaged over $N$ rounds is $N \cdot K / \sum_{i=1}^{N} T_i = K / \overline{T_i}$.

### B. Simulation Setup

In our simulations, we generate 20 random local networks and use them to evaluate the performance of different decomposition schemes. For each network, we generate $N_R=30$ nodes that are randomly distributed between two concentric circles of radii 1 and 0.5 centered on the

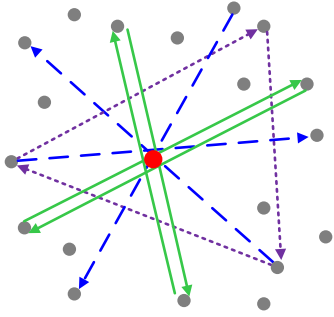

Fig. 5. A random network example.

TABLE 3 SIMULATION RESULTS FOR DIFFERENT SCHEMES UNDER DIFFERENT TRAFFIC VOLUME

| Scheme | K=10 Mean(TS) | K=10 RSD | K=100 Mean(TS) | K=100 RSD | K=1000 Mean(TS) | K=1000 RSD |
|---|---|---|---|---|---|---|
| PNC-9 | 13.7 | 13% | 103.2 | 5% | 978.4 | 0.8% |
| SNC-9 | 16.5 | 5% | 148.3 | 3% | 1449.5 | 1.6% |
| PNC-TWRC | 19.8 | 3% | 183.1 | 2% | 1378.8 | 1.6% |
| Non-NC | 20.0 | 0.0 | 200.0 | 0.0 | 2000.0 | 0.0 |

**TS**: number of time slots;  **RSD**: Relative standard deviation.

relay. Our simulations with different inner radius $r_i$ (from 0.1 to 0.9) yield results that are insensitive to $r_i$ (fluctuations are within 5%). Appendix B gives an explanation. Because of the insensitivity to $r_i$, here we present the results of $r_i = 0.5$ only.

Given a network, we generate $K$ units of traffic, and randomly assign each unit to one of the $F$ node pairs. We generate 20 such random assignments for each network. Since we study 20 networks, this results in 400 experiments for each $K$ (In fact, we tried generating much more than 20 networks and 20 assignments, that is, each data point is the average result of much more than 400 experiments. We found that both the mean and the standard deviation are quite stable when the numbers of random networks and assignments are beyond 20.). In Section VI.C, we present the results averaged over the 400 experiments. We focus on $K$=10, 100, and 1000, representing the low, medium, and heavy traffic scenarios. Fig. 5 shows a random network example with 10 units of traffic. These traffic flows can be scheduled by one atom VI (represented by the normal green arrows), one atom VIII (represented by the dashed blue arrows and one atom III (represented by the purple dotted arrows). Totally ten (3+3+4=10) timeslots are needed by combining these atoms. Except those traffic covered by atom VI, the other traffic flows are non-symmetric, which cannot be scheduled by PNC TWRC. If we only use PNC TWRC, sixteen timeslots are needed.

For the identification of potential flows and PNC atom instances (see Section V.A), we assume the transmission range is one unit and the interference range is 1.78*link length* (a typical setup with wireless path loss exponent of 4 and SIR requirement of 10dB)

## C. Comparison of Different Schemes

The first set of results is presented in Table 3. The numbers of time slots (TS) needed for different schemes when $K$=10, 100, and 1000 are given. Recall that for each $K$, we run 400 experiments. Accordingly, we collect the mean (Mean) and the relative standard deviation (RSD) of the samples.

In Table 3, PNC-9 refers to decomposition using all the nine PNC atom classes; PNC-TWRC refers to decomposition using the PNC TWRC only; Non-NC refers to the traditional scheme in which network coding is not used; and SNC-9 refers to decomposition using the nine corresponding SNC atoms as mentioned in Section III.B.

As shown in Table 3, PNC-9 has significantly better performance than the other schemes under light, medium and large $K$. In particular, for $K = 100$ and $1000$, the transmission efficiency of PNC-9 is about twice that of Non-NC (100% throughput gain). This is because when $K$ is large, the scheduling can be decomposed into numerous PNC atoms to fully exploit the advantages of PNC scheduling.

In [1], it was shown that the throughput gain of PNC in TWRC is 100%. The throughput gains of other PNC atoms can be derived from the data in Table 2: range from 50% to 167% (four of them are exactly 100%). Since the throughput gain of a local network according to Table 3 is also around 100%, we see that the "micro" throughput gains of individual atoms (around 100%) are largely translated to the "macro" throughput gain of the overall network under the PNC decomposition scheme.

## D. Relative Importance of Different Atoms

In this section, we investigate the relative importance of different atoms through extensive simulation experiments under different traffic volumes and network topologies. In particular, we compare the throughputs of different schemes that include different atom classes. A goal is to identify ways to further simplify our decomposition schemes by including only the atoms that are important to good performance. Doing this has two benefits. The first benefit is that the atom identification process will be simplified. As mentioned in Section V.A, the computational complexity of identifying instances of atom class $i$ is $O(F^{M_i}) = O(N_R^{2M_i})$, where $M_i$ is the number of flows in atom class $i$, $F$ is the number of potential flows in the local network, and $N_R$ is the number of peripheral nodes in the local network. If we can exclude the atom classes with large $M_i$, then the computational complexity of the whole identification process can be reduced. The second benefit is that there will be fewer atom instances in the database matrix **D** (fewer variables for the LP), and thus the LP algorithm can be speeded up.

In the following, we summarize our results under a number of key observations followed by the simulation results giving rise to the observations. We begin by re-examining the PNC-9 scheme in more detail for benchmarking purposes.

**Observation 1.** *The performance of the PNC-9 scheme is not sensitive to the network density (the number of peripheral nodes surrounding the relay). It is, however, sensitive to traffic volume K. For small K (e.g., K=10), compared with Non-NC, PNC-9 can achieve throughput gain of*

TABLE 4  SIMULATION RESULTS FOR PNC-9 AND PNC-I UNDER DIFFERENT TRAFFIC VOLUME AND NETWORK DENSITY

| Traffic Volume | Network Density | $N_R=20$ | | $N_R=10$ | | $N_R=6$ | |
|---|---|---|---|---|---|---|---|
| | PNC Scheme | PNC-9 | I | PNC-9 | I | PNC-9 | I |
| $K=10$ | (TS) | 13.7 | 19.8 | 13.9 | 18.1 | 13.7 | 16.0 |
| $K=100$ | (TS) | 103 | 165 | 109 | 136 | 110 | 120 |
| $K=1000$ | (TS) | 980 | 1236 | 1016 | 1116 | 1033 | 1065 |

*Network Density: the density is represented by the number of peripheral nodes ($N_R$).*

TABLE 5  Degradations for Various One-Atom Schemes

| Scheme | I | II | III | IV | V | VI | VII | VIII | IX | PNC-6 |
|---|---|---|---|---|---|---|---|---|---|---|
| $K=100$ $N_R=10$ | 21% | 14% | 39% | 35% | 18% | 44% | 44% | 42% | 45% | 32% |

*PNC-6: the PNC scheme using atoms III, IV, VI, VII, VIII and I.*

TABLE 6  Degradations for PNC-I&V and PNC-I&II&V

| Scheme | PNC-I&V | | | | PNC-I&II&V | | | |
|---|---|---|---|---|---|---|---|---|
| Density | $N_R=30$ | $N_R=20$ | $N_R=10$ | $N_R=6$ | $N_R=30$ | $N_R=20$ | $N_R=10$ | $N_R=6$ |
| $K=10$ | 5% | 6% | 7% | 7% | 2% | 2% | 1% | 1% |
| $K=100$ | 3% | 3% | 4% | 4% | 2% | 2% | 1% | 1% |
| $K=1000$ | 2% | 2% | 2% | 2% | 1% | 2% | 1% | 0% |

*around 46% regardless of the network density. For medium and large K, (e.g., K=100 or 1000), the gain is around 100% regardless of the network density.*

Observation 1 is supported by our second set of simulation results presented in Table 4. Here we use the number of peripheral nodes $N_R$ to represent the network density. In our simulations, for each $K$, we set $N_R$ = 30, 20, 10 and 6 to represent the dense to sparse networks. We run 400 experiments for each pair of $K$ and $N_R$.

Since we have already presented the result of $N_R=30$ in Table 3, we only present the results of $N_R=20$, 10 and 6 in Table 4. We omit the RSDs in Table 4 because they are similar to those presented in Table 3 for a given $K$. As shown in Tables 3 and 4, for a given $K$, the numbers of timeslots consumed by PNC-9 under different network densities are about the same (the variation is within 5% ).

The result can be understood intuitively as follows. In general, when the network is sparse (small $N_R$), there are few potential flows (the number of which is bounded by $N_R(N_R -1)$). Then, each flow will have less chance to find matching flows to form an atom instance. However, armed with nine possible atom classes, PNC-9 still allows the limited combinations of flows to be matched to some of the atom classes to form atom instances. That is, with PNC-9, any given flow is highly likely to be involved in at least one atom instance.

Table 4 also lists the performance results of PNC-I, in which only atom class I, the TWRC, is used in the decomposition (i.e., PNC-I is PNC-TWRC). For simplicity, hereafter we will represent different decomposition schemes by the roman numerals of the atom classes being used. For example, PNC-I&V is the decomposition scheme that makes use of only atom class I and atom class V.

From Table 4, we see that there are still appreciable performance gaps between PNC-I and PNC-9 except under high traffic volume ($K=1000$) and sparse network ($N_R=6$). Thus, using only atom I is not good enough in general.

The natural question that follows is whether any of the other eight atoms will be good enough when they are applied individually in the PNC decomposition scheme. The answer turns out to be negative, as detailed below.

**Observation 2.** *Using just one atom class from the nine classes studied in the PNC decomposition scheme does not yield good performance. Specifically, the performance gap between each of PNC-I, PNC-II, ... , PNC-IX and PNC-9 is large.*

Observation 2 is supported by our third set of simulation results. Before proceeding further, let us clarify how we measure the performance degradation of a scheme $i$ with respect to the benchmark scheme PNC-9. For a given $N_R$ and $K$, we conduct 400 experiments. For the $n_{\text{th}}$ experiment, we calculate the throughput degradation of scheme $i$ (in percentage) with respect to PNC-9 by

$$D_{i,n} = \left( \frac{1}{T_{PNC-9,n}} - \frac{1}{T_{i,n}} \right) \bigg/ \frac{1}{T_{PNC-9,n}} = \frac{T_{i,n} - T_{PNC-9,n}}{T_{i,n}} \times 100\%$$

where $T_{i,n}$ is the number of time slots consumed by scheme $i$, and $T_{PNC-9,n}$ is the number of time slots consumed by PNC-9. Then, we average $D_{i,n}$ over the 400 experiments by $D_i = \sum_n D_{i,n} / 400$.

As a convenient benchmark, let us say that a scheme $i$ is "good enough" (or "nearly as good as PNC-9") if $D_i$ is less than 10% under all pairs of $N_R$ and $K$ being studied. Note that the average throughput degradation is not the only possible metric to evaluate a scheme; later in Observation 4 we will further evaluate various schemes by the tail distribution of the degradation.

Table 5 presents the average throughput degradation $D_i$ for various one-atom schemes for $N_R=10$ and $K=100$. As can be seen, the performance gap ranges from 14% to 45 %. None of PNC-I, PNC-II, ... , PNC-IX is "good enough".

From Table 5, we also notice that, PNC-I, PNC-V and PNC-IX have better performance than other one-atom schemes. Although not presented here, PNC-I, PNC-V and PNC-IX are also the top three performers for other $N_R$ and $K$. This suggests that atoms I, II & V are the three most important atom classes in terms of providing good performance. For further validation, we perform additional experiments in which atoms I, II & V are excluded, namely only the remaining six other atoms are used. We refer to this scheme as PNC-6. The performance gap between PNC-6 and PNC-9 is considerable (e.g., in Table 5, the gap is ~32% under $N_R=10$ and $K=100$).

On the other hand, if we apply atoms I, II & V in combination and exclude the other six atoms, as shown in Table 6, under various $N_R$ and $K$, the performance gaps between PNC-I&II&V and PNC-9 are negligible (the maximum gap is ~2%).

Next, we explore whether using just two atom classes is good enough. Here, we only focus on the $C_3^2 = 3$ combinations of two atoms among atoms II & V.

**Observation 3.** *PNC-I&V is nearly as good as PNC-9. The maximum gap between PNC-9 and*

*PNC-I&V is 7% under sparse network ($N_R$=6) and low traffic volume (K=10).*

Observation 3 is supported by our fourth set of simulation results in Table 6. As shown in Table 6, PNC-I&V (the use of only atoms I and V) is nearly as good as PNC-9. The performance gap between PNC-9 and PNC-I&V is at most 7%, which occurs under sparse network ($N_R$=6) and low traffic volume (K=10). The other two combinations, PNC-I&IX and PNC-V&IX, incur more than 25% and 19% degradations, respectively, under $N_R$ =30 and K=10, and $N_R$ =6 and K=1000 (results not shown in Table 6 to conserve space).

There are mainly five reasons[3] why the combination of atoms I and V can perform as well as the combination of nine atom classes in PNC decomposition:

a) Each of atom I or atom V can achieve 100% throughput gain compared to the non-NC scheme for the traffic flows they support. The overall throughput gain of applying PNC-9 in a local network according to Tables 3 and 4 is in the ballpark of 100% only and not beyond. Thus, if most traffic units can be matched to atom instances of class I or V, the throughput gain for the overall network can also be close to 100%.

b) For a typical local network, instances of atoms I and V are more prevalent. With reference to Fig. 2, we see that both atoms I and V only have two flows, and the CI requirements between the two flows are simple (compared to other atoms). By contrast, although the throughput gain of atom IX is 140%, instances of it are hard to find because it requires eight flows being matched to complex CI requirements.

c) Atoms I and V deliver symmetric traffic flows and asymmetric traffic flows respectively. We can divide our nine atoms into two groups according to whether their traffic flows are symmetric or not -- by symmetric flows, we mean when there is a flow from A to B, there is a corresponding flow from B to A. Including atoms I and V ensures that there is one representative from each group, so that we can cater to both symmetric and asymmetric traffic. Note that atom I is the smallest PNC block with symmetric flows. We can schedule multiple atom-I instances to satisfy the multiple symmetric traffic flows in other atoms with symmetric flows such as atom VI and atom IX. Meanwhile, atom V is one of the two smallest PNC blocks with asymmetric flows[4]. The two cross flows in atom V are embedded in other atoms with asymmetric flows such as atom VII and atom VIII. Therefore, we can use multiple atom-V instances to satisfy the traffic flows in these atoms.

d) Atoms IV and VII can be replaced by atoms I and V without throughput degradation.

---

[3] The first two reasons also apply to atom II.
[4] The other one is atom II.

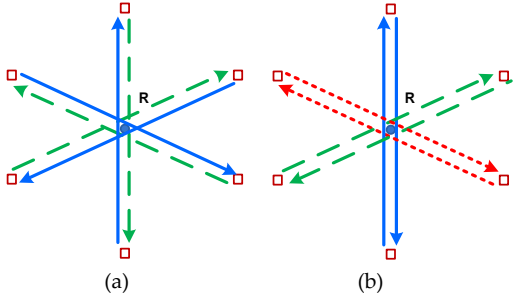

| TABLE 7 | | | | TAIL PROBABILITY P(D>10%) | | | | |
|---|---|---|---|---|---|---|---|---|
| Scheme | PNC-I&V | | | | PNC-I&II&V | | | |
| Density | $N_R=30$ | $N_R=20$ | $N_R=10$ | $N_R=6$ | $N_R=30$ | $N_R=20$ | $N_R=10$ | $N_R=6$ |
| $K=10$ | 17% | 25% | 34% | 39% | 5% | 3% | 3% | 5% |
| $K=100$ | 1% | 1% | 5% | 8% | 0% | 1% | 1% | 2% |
| $K=1000$ | 0% | 0% | 1% | 1% | 0% | 0% | 1% | 2% |

(a)   (b)

Fig. 6. Demonstration of Molecule VIII.

With reference to the structure of atom IV in Fig. 2(d), we see that atom I can be used to carry the symmetric flows between nodes B and C, and the regular two-hop non-NC scheme can be used to deliver the traffic from A to B. This consumes a total of four timeslots, the same as in atom IV. For atom VII in Fig. 2(g), atom V can be used to carry the two cross flows from B to E and from C to F, and then the two-hop non-NC scheme can be used to deliver the traffic from A to D. This consumes four timeslots, the same as in atom VII.

e)   Atoms III and VIII can be replaced by the combination of atoms I and V with only small throughput degradation. We first examine atom VIII in Fig. 2(h). If we use atom V to carry the two cross flows from E to B and from F to C, and the two-hop non-NC scheme to carry the traffic from A to D, a total of four time slots is needed. This is one more time slot than is needed by the transmission pattern of atom VIII. When the total traffic volume is small, penalty of this type may cause a significant performance gap percentage-wise. However, when the total traffic volume is large, the percentage performance penalty will be low, as elaborated below.

Once we identify an atom-VIII instance, in the same hexagonal topology, we can also identify another atom-VIII instance with flows in the opposite directions as shown in Fig. 6(a). This is because if the topology satisfies the CI requirement of one instance, it must satisfy the CI requirement of the other instance. We refer to such a pair of atom-VIII instances as "molecule VIII". A molecule VIII requires six time slots when it is scheduled according to the transmission pattern of the two underlying atom VIII's. We note that a molecule VIII can also be decomposed into three atom I's (as shown in Fig. 6(b)). Scheduling according to the transmission patterns of the three atom I's also require six slots, no more than scheduling with two atom VIII's. Thus, eliminating atom VIII will not cause penalty (as long as atom I is included) when the traffic volume $K$ is large relative to the number of potential flows $F$ in the local network. This is because most potential flows will have traffic and it is likely that all flows in the topological structure of molecule VIII will have traffic.

Similarly, although an individual atom III cannot be replaced by other atoms, for each atom-III instance there must a counterpart atom-III instance with flows in the opposite direction in the triangular topology. Such a "molecule III" can be decomposed into three atom-I instances. The total timeslots needed is 3×2=6, which is even smaller than scheduling the molecule III using the transmission patterns of the two underlying atom III's (2×4=8). Again, when $K$ is large relative to $F$, most pairs of atom-III's can be scheduled as molecule III's.

Overall, we can say that substituting atoms III and VIII with atoms I and V will cause negligible performance degradation on average, especially when $K$ is large relative to $F$.

Overall, the conclusion from observations 2 and 3 are as follows:

**Conclusion 1.** *For good throughput performance averaged over different network topologies, PNC-I&V is a good substitute for PNC-9.*

So far, we have focused on the performance of PNC-I&V averaged over different simulation runs with different network topologies and different traffic distributions across the potential flows. We next investigate the tail distribution of the degradation $D_{i,n}$. Here, we are interested in a different performance metric, $\gamma = \Pr_{Di}(D_{i,n} > 10\%)$. In other words, we define degradation of more than 10% as being unacceptable and would like to evaluate the chance for that. As a rule of thumb, let us define $\gamma \leq 5\%$ to mean the performance target is met most of the time and is acceptable.

**Observation 4.** *For PNC-I&V, under low traffic volume ($K=10$), the performance target of $D_{PNC\text{-}I\&V, n} \leq 10\%$ is often not met for specific network topologies and traffic distributions over potential flows; for PNC-I&II&V, the performance target of $D_{PNC\text{-}I\&II\&V, n} \leq 10\%$ is always met regardless of $N_R$ and $K$.*

Observation 4 is supported by our fifth set of simulation results presented in Table 7. Table 7 lists $\gamma$ for PNC-I&V and PNC-I&II&V under various $N_R$ and $K$. As can be seen, for PNC-I&V, under low traffic volume ($K=10$), $\gamma > 5\%$ regardless of $N_R$; under medium traffic volume ($K=100$) and sparse networks ($N_R=6$), $\gamma$ is 8%, also larger than 5%. This means that there are still many network topologies and traffic distributions in which the degradation is not acceptable, although under high traffic volume ($K=1000$), the performance target is met most of the time. On the other hand, for PNC-I&II&V, $\gamma \leq 5\%$ regardless of $N_R$ and $K$.

In the preceding reason c) for Observation 3, we mentioned that atom V is one of two smallest PNC blocks with asymmetric flows. The other one turns out to be atom II. Also, atom II cannot be substituted by atoms I and V. Hence, when the traffic volume is small, the penalty of excluding atom II may cause a significant degradation percentage-wise, especially under a sparse network (in which there are only a limited number of atom instances).

The combined intuition from the discussion of the two preceding paragraphs is as follows. Let us refer to atom instances whose flows all have traffic as *fully-occupied atom instances*. When the number of fully-occupied atom instances is small, the degradation of PNC-I&V may not be acceptable; we need to use PNC-I&II&V instead.

**Conclusion 2.** *For good throughput performance in different network topologies and under different traffic volume scenario, PNC-I&II&V is a good substitute for PNC-9. An implication is that atoms I, II, V is a useful subset of atoms that are most important to good performance.*

## VII. IMPLEMENTATION ISSUES

In this section, we investigate how our PNC decomposition scheme can be deployed in practice. Specifically, we address a number of practical implementation issues.

A key issue is who will perform the computations as described in the previous section. We assume that the relay is responsible for doing so.

First, the relay needs to find out the structure of the local topology: (i) the peripheral nodes in the transmission range of the relay; (ii) for each peripheral node, the other peripheral nodes that are within its transmission range and interference range. Distributed algorithms for topology discovery have been well studied (see, for example, the power-exchange topology discovery algorithm in [20] and [21], and the topology discovery algorithm for hybrid wireless networks in [22]). We will therefore not delve into it here. Given the local topology information, the relay can then identify the atom instances to construct the database as described in Section V.A.

Next, the relay needs to find out the traffic demands between the peripheral nodes. In particular, we need a MAC protocol with which the "source" peripheral nodes could communicate their demands to the relay. Based on these demands, the relay then schedules and orchestrates the transmissions within the local network.

In Section VII.A, we present a MAC protocol for the above purpose. An issue that arises is whether the MAC protocol will incur large overhead. In Section VII.B, we present a frame design for our protocol and show that the overhead is small. In Section VII.C, we present the simulation results of various schemes under distributed coordination protocol.

### A. MAC Design Principles

Our proposed protocol is similar to the Point Coordination Function (PCF) in IEEE 802.11. In the following, we provide details and variations of our design:

1. The relay node serves the role of the point coordinator (PC) in our PCF [IEEE 802.11 standard]. The local network around the relay is analogous to the basic service set (BSS) in IEEE 802.11.

2. Our PCF adopts a two-step multi-polling scheme: the first polling frame (referred to as the

| Bytes: 2 | 6 | $N_R/8$ | 1 | 4 |
|---|---|---|---|---|
| Frame Control | BSSID | Polling BitMap | Window Size | FCS |

(a) Multi-poll request frame.

| Bytes: 2 | 6 | 2 x W | | 4 |
|---|---|---|---|---|
| Frame Control | BSSID | DID (2) | ….. | FCS |

(b) Multi-poll demand frame.

| Bytes: 2 | 6 | Polling Control(2+2W) x $N_R$ | | | | 4 |
|---|---|---|---|---|---|---|
| | | | 2 x W | | | |
| Frame Control | BSSID | SID (2) | Start Time (1) | Atom Role (1) | ….. ….. | FCS |

(c) Multi-poll assignment frame

Fig. 7. Frame formats.

TABLE 8  THROUGHPUT GAINS OF PNC-9 RELATIVE TO NON-NC AND PNC-I

| Density | $N_R=30$ | | $N_R=20$ | | $N_R=10$ | | $N_R=6$ | |
|---|---|---|---|---|---|---|---|---|
| Scheme | Non-NC | I | Non-NC | I | Non-NC | I | Non-NC | I |
| $W=1$ | 85% | 79% | 75% | 70% | 53% | 51% | 46% | 45% |
| $W=2$ | 92% | 83% | 82% | 75% | 61% | 58% | 55% | 53% |
| $W=3$ | 96% | 84% | 86% | 78% | 65% | 62% | 59% | 58% |
| $W=4$ | 98% | 83% | 87% | 78% | 67% | 60% | 61% | 60% |

multi-poll request frame) is broadcast by the relay to collect the traffic demand information from the peripheral nodes[5]. Then each polled peripheral node replies with a frame (referred to as the multi-poll demand frame) containing information on its pending packets, including the peripheral nodes to which they are destined. The maximum number of pending packets to be reported is limited to W. Based on the collected traffic demands, the relay solves LP (2) to get the optimal scheduling. Instead of LP (2), simpler heuristic algorithms could also be used. We will consider a particular simple heuristic algorithm later. The second polling fame (referred to as the multi-poll assign frame) is then broadcast by the relay. The purpose of this frame is to inform the peripheral nodes of their assigned transmit time slots and the particular packets they should transmit in the assigned time slots. The above two-step multi-polling is followed by a sequence of transmissions (referred to as a *round*) by the peripheral nodes and the relay.

3. We adopt an end-to-end acknowledgement (ACK) mechanism. Specifically, after the sequence of transmissions by all nodes, each destination peripheral node then sends the relay a report containing ACKs for the packets that have been received correctly. Then the relay summarizes the ACKs and broadcasts the summarized information to the source peripheral nodes. The lost packet will be rescheduled in the next round. Note that in the performance evaluation of this paper, we do not stimulate the ACK mechanism and assume the reliable transmission in the physical layer. The detailed ARQ design for PNC atoms will be investigated in our future work.

*B. Frame Format and Overhead*

In our PCF model, as mentioned above, there are three types of frames that need to be designed. We propose the frame formats as shown in Fig. 7.

Fig. 7(a) shows the frame format of multi-poll request, whose length varies with the number of peripheral nodes. Other than the new Polling Bitmap field and Window Size field, the other fields follow the 802.11 standard. Each bit of the Polling Bitmap field represents a peripheral node. If a peripheral node is scheduled to transmit in a round, its bit is set to 1. Thus, if there are totally $N_R$ peripheral nodes, $N_R/8$ bytes are needed. The Window Size field indicates the maximum number of packets W that each node can transmit in the round. Overall, the multiple-poll request frame requires $13+N_R/8$ bytes.

---

[5] Here, as assumed in the previous sections, only the traffic routed through the relay is considered.

Fig. 7(b) shows the frame format of multi-poll demand. In this frame, we introduce a new field called the destination identifier (DID). The DID is used to identify a station. In particular, each DID field indicates the "destination" peripheral node (in the local network) of one packet. If a maximum of W packets can be scheduled per node, we need W DID fields. Overall, the multiple-poll demand frame requires $(12+2W)$ bytes. For $N_R$ peripheral nodes, a total of $(12+2W)N_R$ bytes are required.

Fig. 7(c) shows the frame format of multi-poll assignment, whose length depends on the number of peripheral nodes being polled and the window size. The Polling Control field consists of three subfields: source identifier (SID), Start Time and Atom Role. The SID identifies a polled peripheral station that has source packets to be scheduled. Then for each of the station's packets (up to a maximum of W packets) to be scheduled, there is a Start Time and an Atom Role, as indicated by their respective fields. Since each packet is assigned to an atom instance, the Start Time indicates the first time slot of the assigned instance; the Atom Role indicates the role played by the node in this instance (i.e., which node it is within the atom class structure). Specifically, the first four bits of the Atom Role field identifies the atom class, and the last four bits identifies the specific node within the class to which the polled node corresponds. Overall, the Start Time and Atom Role tells the node when it should transmit its packet.

Going into further details, consider atom VI as an example (see Fig. 2(f)). In the first time slot, node A transmits, node B receives. Therefore each peripheral node needs to maintain a template for each atom class to indicate the detailed actions of different nodes[6]. During the execution of the transmission pattern, the source node does not need to know who its partners are (i.e., the other source nodes that belong to the same atom); that is, each source node can independently complete its own task by following the instruction indicated by the Atom Role. For example, if a node is to act as node A of atom VI, in the first time slot (the Start Time) it transmits its own packet; in the second time slot it receives (the simultaneous transmitting packets) and decodes them into XOR form; in the third time slot it receives the broadcast XOR packet from the relay and extracts its desired packet by doing XOR. Overall, the polling control field requires

---

[6] Note that in our atoms, besides the relay, only the source nodes will participate in the transmission (including transmitting its own packet and relaying others' packets).

$(2+2W)N_R$ bytes and the whole frame requires $12+(2+2W)N_R$ bytes.

Given the above frame formats, the extra overhead ($H$) due to our distributed coordination protocol is $H=(13+N_R/8)+(12+2W)N_R+(12+(2+2W)N_R)$ bytes[7]. The overhead ($\overline{H}$) per packet when all nodes have $W$ packets to transmit in each round is $\overline{H}=H/(N_R \cdot W) \approx 4+14/W+25/(N_R \cdot W)$. From the expression we can see that with the increase of traffic volume, namely large $W$ and $N_R$, $\overline{H}$ will be small (approximately 4 bytes). Even considering the largest $\overline{H}$ case in our simulations (to be presented in Part C) when $N_R=6$ and $W=1$, $\overline{H}$ is only 22 bytes, which is a normal length for the control frames in 802.11. Overall, the extra cost specific to our PNC scheduling is small, especially in view of its significant throughput gain.

*C. Simulation*

In Section VI, we assumed each packet is equally likely to be on each of the potential flows in the local network. For our simulations here, we consider a saturated network in which all peripheral nodes always have $W$ or more packets to send. However, as described above, at most $W$ packets from each node will be scheduled in each round. We assume each packet from a node is equally likely to be on the potential flows coming out of the node. Since, in general, there may be different numbers of potential flows coming out of different nodes, this may not give rise to a uniform traffic distribution across all the potential flows in the local network, as was assumed in Section VI. For our simulations here, in each round the total traffic volume $K$ in the local network is $WN_R$. Other simulation setups such as the generation of the random network and the identification of the atom instances are the same as in Section VI.B[8].

Table 8 presents our seventh set of simulation results obtained under various $N_R$ and $W$ (from 1 to 4). We compare PNC-9 with Non-NC and PNC-I and list the throughput gains provided by PNC-9 relative to them. When $K$ is around 10, that is, $N_R=10$ and $W=1$ (or $N_R=6$ and $W=2$), the throughput gain by PNC-9 relative to Non-NC is around 50%; when $K$ is around 100, that is, $N_R=30$ and $W=3$ (or 4), the throughput gain by PNC-9 relative to Non-NC is around 100%. Both gains are close to the gains from the data obtained under similar $K$ in Observation 1 of Section VI (see Table 3). In addition, the throughput gain relative to PNC-I is also significant, implying that applying only atom I is still not good enough. Since the extra distributed coordination overhead is small, under the distributed coordination scheme, PNC-9 can achieve performance as good as that under the centralized scheme in Section VI.

---

[7] Here we do not consider typical overheads such as preambles, short interframe space (SIFS) and ACKs, because such overheads are needed in any distributed MAC protocol, not just ours.
[8] We do not simulate the ACK mechanism here.

TABLE 9    SIMULATION RESULTS OF PNC-I&II&V

|         | Degradation relative to PNC-9 | | | | Tail Probability $P_{(D>10\%)}$ | | | |
|---------|-------|-------|-------|-------|-------|-------|-------|-------|
| Density | $N_R=30$ | $N_R=20$ | $N_R=10$ | $N_R=6$ | $N_R=30$ | $N_R=20$ | $N_R=10$ | $N_R=6$ |
| $W=1$ | 2% | 2% | 1% | 1% | 1% | 2% | 3% | 4% |
| $W=2$ | 2% | 3% | 2% | 1% | 0% | 1% | 2% | 5% |
| $W=3$ | 2% | 2% | 1% | 1% | 0% | 1% | 1% | 4% |
| $W=4$ | 2% | 3% | 1% | 1% | 0% | 1% | 1% | 3% |

TABLE 10    DEGRADATION FOR NON-LP ALGORITHM

| Scheme | PNC-9 | | | | PNC-I&II&V | | | |
|---------|-------|-------|-------|-------|-------|-------|-------|-------|
| Density | $N_R=30$ | $N_R=20$ | $N_R=10$ | $N_R=6$ | $N_R=30$ | $N_R=20$ | $N_R=10$ | $N_R=6$ |
| $W=1$ | 11% | 10% | 6% | 4% | 8% | 7% | 3% | 2% |
| $W=2$ | 10% | 8% | 7% | 5% | 7% | 6% | 4% | 2% |
| $W=3$ | 10% | 8% | 6% | 7% | 6% | 6% | 3% | 2% |
| $W=4$ | 11% | 10% | 5% | 7% | 7% | 6% | 3% | 3% |

Table 9 lists the performance results of PNC-I&II&V under various $N_R$ and $W$. We first consider the average throughput degradation by PNC-I&II&V relative to PNC-9. We can see that the degradation is similar to that under the centralized scheme in Section VI. For $\gamma = \Pr(D_{PNC-I\&II\&V, n} > 10\%)$, we see that $\gamma \leq 5\%$ regardless of $N_R$ and $W$. Hence, PNC-I&II&V is a good substitute for PNC-9 under the distributed scheme.

Let us now consider replacing the LP with a simple heuristic algorithm to accelerate the computation process. Here, we propose a simple greedy algorithm that gives higher priority to atom instances that are more efficient. We only schedule the "fully-occupied atom instances" (see the last paragraph of Section VI for definition). Specifically, we will first construct a reduced database matrix **D** based on only the node pairs that have traffic between them. The details on the construction of **D** can be found in [23]. Then, instead of running the LP on **D**, we sort the atom instances in **D** according to their throughput efficiency (i.e., the ratio of the number of time slots required by the PNC atom to the corresponding Non-NC atom) to obtain an ordered reduced matrix **D**. From Table 2, the priorities of the atoms can be ordered as follows: atom VI, atom IX, atom I, atom II, atom V, atom VIII, atom III, atom IV and atom VII. In the scheduling, we start by scheduling the first instance in one or more times until there is no more packet on one of the traffic flows in this instance. Then, we move on to the next "fully-occupied atom instances" and schedule this instance similarly. We continue until all packets have been scheduled.

For typical sorting algorithms, the computational complexity is always less than the LP. For example, merge sort is an $O(n \log n)$ sorting algorithm. Furthermore, the sorting can be done during the identification process. Specifically, we identify the most efficient atom class first, then the second most efficient, and so on. As a result, we do not need to sort them in each round.

Without solving the LP, the scheduling may not be optimal. However, our simulation results in Table 10 indicate that for PNC-9, the degradation incurred by the simple sorting algorithm ranges from 4% to 11% only; furthermore, for PNC-I&II&V, the degradation ranges from 2% to 8% only.

## VIII. CONCLUSION

This paper has introduced the concept of PNC atoms as building blocks of PNC networks. This paper is a first attempt to apply PNC to larger networks in a systematic way. We have identified nine PNC atoms, with TWRC being one of them. We investigated a PNC atom decomposition framework to solve the scheduling problem in PNC networks.

Our performance evaluation results indicate that decomposition based on the nine PNC atoms can yield throughput gain of about 100% compared with the traditional multi-hop (non-NC) scheduling (40% throughput gain compared with decomposition based on the TWRC PNC atom alone).

We further analyzed the relative importance of the nine atoms. We found that among the nine atoms, atoms I, II and V are particularly instrumental to good throughput. In particular, decomposition based on atoms I, II and V yields throughput performance that is almost as good as decomposition based on all nine atoms.

We have also designed a low-overhead MAC protocol to coordinate the transmissions of different nodes according to the scheduling results of PNC decomposition.

In this paper we have proposed and investigated a PNC decomposition framework for application in local networks with only one relay. Going forward, a particular important future direction is how the concept of PNC decomposition can be extended and applied in large networks consisting of multiple relays.

## APPENDIX A. OPTIMALITY OF TRANSMISSION PATTERNS

In this appendix, we show that the nine PNC atom classes in Fig. 2 are optimal in the sense that their transmission patterns consume the minimum numbers of time slots given their topological structures. Recall that in designing the transmission pattern, we can only rely on the connectivity and interference-free relationships given in the CI-graph (Section III.A). This assumption is also made in the proofs presented in this appendix. For example, when we say two nodes cannot hear each other, we mean there is no C-edge between the two nodes in the CI-graph, and therefore the transmission pattern cannot assume that they can hear each other.

In the following, by "the information about packet $S_i$", we mean either the native packet $S_i$ or a mixed packet of the form $S_i \oplus \textit{some other packets}$. Essentially, each reception by a node gives the node a linear XOR equation. A node does not have the information about packet $S_i$ means none of the equations it has contains $S_i$.

We note that the assumption that destination $D_i$ cannot directly receive from its source node $S_i$ (Assumption 2) does not imply that $D_i$ cannot obtain any information about its desired packet $s_i$ during the uplink phase. The information about $s_i$ can be propagated by other source nodes to $D_i$ in the time slots subsequent to the time slot in which node $S_i$ transmits information about packet $s_i$. That is, $D_i$ could potentially obtain the information about $s_i$ through overhearing other source nodes.

We put forth three propositions below to prove the optimality of the transmission patterns of the nine PNC atoms. For simplicity, the three propositions are based on the assumption that only the relay transmits in the downlink phase. However, we add a remark after each proof explaining that even if we allow peripheral nodes to transmit together with the relay in the downlink phase, such schemes will not be more optimal than the scheme in which only the relay transmits in the downlink phase.

**Proposition A.1:** The number of time slots needed for a PNC atom to deliver one packet for each flow is lower bounded by $\lceil N/2 \rceil + 1$, where $N$ is the number of source nodes in the atom.

**Proof:** Recall that in Section III.A, we divide the transmission process into uplink phase and downlink phase. For practicality and simplicity, we assume that relay R can receive from at most two nodes at one time (in Assumption 4). Therefore, we need at least $\lceil N/2 \rceil$ time slots for the uplink phase. For the downlink phase, at least one time slot is needed. To see this, consider a source node $S_i$ that transmits in the last of the $\lceil N/2 \rceil$ uplink time slots. The corresponding destination node $D_i$ cannot hear the packet $s_i$ (Assumption 2). Meanwhile, there are no uplink time slots left for information about packet $s_i$ to propagate through the transmissions of other source nodes to $D_i$. The assistance of the relay R is therefore needed to forward information related to packet $s_i$. **Q.E.D.**

**Remark**: It is obvious that Proposition A.1 is valid regardless of whether peripheral nodes can transmit in the downlink phase.

**Atoms that require $\lceil N/2 \rceil + 1$ time slots:** Atoms I, II, V, VI, and VIII in Fig. 2 consume $\lceil N/2 \rceil + 1$ time slots, achieving the above lower bound. Thus, for these atoms, there are no better transmission patterns than the one we give.

**Atoms that require $\lceil N/2 \rceil + 2$ time slots:** Atoms III, IV, VII and IX in Fig. 2 consume $\lceil N/2 \rceil + 2$ time slots. However, their topological structures are such that this is necessarily so.

That is, the achievable lower bound for these atoms must be no less than $\lceil N/2 \rceil + 2$. This is captured in Propositions A.2 and A.3 below.

**Proposition A.2:** Consider a PNC atom with $N$ source nodes. Suppose that 1) each and every source, besides not being heard by its own destination, cannot be overheard by at least one other destination node during the uplink phase; and that 2) the topological structure of the atom is such that it is not possible for each and every destination to obtain any information about its desired packet at the end of the uplink phase if the uplink phase has only $\lceil N/2 \rceil$ time slots. Then $\lceil N/2 \rceil + 2$ time slots are needed to deliver one packet for each flow.

**Proof:** As explained in the proof of Proposition A.1, the number of time slots required for the uplink phase is lower-bounded by $\lceil N/2 \rceil$, and in each time slot at most two sources transmit. This means that at least one source must transmit for the first time in the last of the $\lceil N/2 \rceil$ time slots. Let this source node be denoted by $S_i$. By <u>supposition 1)</u>, this source node, besides not being heard by its destination $D_i$, cannot be overheard by another destination node, say $D_j$. Let the desired packet of $D_j$ be denoted by $S_j$. Both $D_i$ and $D_j$ do not have any information about packet $S_i$ at the end of the uplink phase, since they cannot directly overhear node $S_i$, and node $S_i$ transmits in the last uplink time slot. By <u>supposition 2)</u>, $D_j$ does not have information about its desired packet $S_j$ either by the end of the uplink phase. Now, suppose that relay R uses only one time slot for the downlink phase. The above implies that the relay must transmit a network-coded packet in the form of $S_i \oplus S_j$ or $S_i \oplus S_j \oplus$ *some other packets* in order that nodes $D_i$ and $D_j$ could obtain some information about their desired packets $S_i$ and $S_j$, respectively. However, since node $D_j$ does not have any information about $S_i$ from the uplink phase receptions, it cannot cancel out $S_i$ from the network-coded packet transmitted by R to obtain $S_j$. Thus, at least two time slots are needed for the downlink phase given suppositions 1) and 2). **Q.E.D.**

Suppositions 1) and 2) are true in atoms III, IV, and VII. Thus, they need at least $\lceil N/2 \rceil + 2$ time slots. For illustration, consider atom VII, in which $N = 3$. It is easy to verify from the CI-graph in Fig. 2(f) that supposition 1) applies. As for supposition 2), note that $\lceil N/2 \rceil = \lceil 3/2 \rceil = 2$. Let us consider the source-destination pair (A, D). Not including R, node D is three hops away from node A. Thus, given that there are only two time slots in the uplink phase, it is impossible for the information about packet *A* to propagate to node D by the end of the uplink phase. Similar arguments apply for pairs (B, E) and (C, F).

**Remark:** Now, let us see if, rather than two time slots, we can use only one time slot for the downlink phase if we allow one peripheral node to transmit together with the relay in that time slot (recall that we assume at most two nodes transmit simultaneously). If this is possible, then we can reduce the lower bound back to $\lceil N/2 \rceil+1$. We change <u>supposition 2)</u> in Proposition A.2 to <u>supposition 3)</u>: the topological structure of the atom is such that it is not possible for each and every destination to obtain any information about its desired packet at the end of the downlink phase if the uplink phase plus downlink phase has only $\lceil N/2 \rceil+1$ time slots. Only if suppositions 1) and 3) are true in an atom, then at least $\lceil N/2 \rceil+2$ time slots are needed to deliver one packet for each flow of this atom.

For atoms III and IV, since the peripheral nodes are out of each other's transmission range (i.e., they can only be heard by the relay and not by any other peripheral node), it is not meaningful to allow the relay and the peripheral node transmit simultaneously. Suppositions 1) and 3) are certainly true in atoms III and IV.

For atom VII, we can perform an exhaustive exploration of all possible cases to show that it is still impossible to deliver its traffic in $\lceil N/2 \rceil+1=3$ time slots. For illustration, let us consider two of the cases. Certainly, out of the three time slots, two will be needed for the uplink phase and one for the downlink phase if we limit ourselves to three time slots. Since each source-destination pair is three hops away, only the packet that is transmitted in the first time slot can arrive its destination through the peripheral nodes (not through the relay), that is, the packet first transmitted in the second time slot can only arrive its destination through the relay.

1) Nodes B and C transmit in the first time slot and node A transmits in the second time slot. Neither E (destination of B) nor F (destination of C) can overhear A. In the downlink phase, if there is only one downlink time slot, R will have no choice but to transmit $A$ or $A \oplus \text{some other packets}$ to destination D. Now, node E desires packet $B$. If node E wants to get $B$ through the relay, it must eliminate packet $A$. It means node C must help forwarding $A$ to E in the third time slot. If node E wants to get $B$ through the peripheral nodes (B→A→C→E), node C also must transmit in the third time slot. (The route B→F→D→E is impossible because D has to listen from the relay in the third time slot.) By symmetry of nodes B and C, we conclude that both nodes C and B must transmit in this third time slot for destinations E and F to get their desired packets. However, only one peripheral node can transmit in the third time slot together with the relay. So it is impossible that all the destinations get their desired packets within three time slots.

2) Node A transmits in the first time slot, and nodes B and C transmit in the second time slot. Other than the relay, node D (destination of A) can only get the information about packet *A* either from E or from F. However, both E and F are destinations who cannot get their desired packets within uplink phase; they have to receive from the relay in the downlink phase. Thus, it can be seen that relay R will have no choice but to transmit $A \oplus B \oplus C$ in the third time slot. Since node D cannot overhear B and C in the second time slot, it cannot eliminate B and C to extract *A*.

The remaining atom is atom IX. Unfortunately, supposition 2) in Proposition A.2 does not apply to atom IX. In Proposition A.3 below, we show that the topological structure of atom IX is such that it still requires at least $\lceil N/2 \rceil + 2$ time slots.

**Proposition A.3:** The topological structure of atom IX is such that a minimum of five time slots is needed for the delivery of one packet for each flow.

**Proof:** In atom IX, $N = 6$ and $\lceil N/2 \rceil = 3$. Suppose that the uplink phase uses three time slots (note: this is the minimum number of uplink time slots needed because we assume at most two sources can transmit in the same time slot – Assumption 4) and the downlink phase uses just one time slot for a total of four time slots. In the following, we show that no matter how we schedule the transmissions in atom IX, it is not possible to use only four time slots.

Not including R, the destination of each source is three hops away from the source. This means that it will be possible for the information of a source packet to propagate to its destination in the three uplink time slots only if the corresponding source node transmits in the first time slot. However, only two of the sources can transmit in the first time slot (Assumption 4). Among the six sources, two (denoted by $S_{11}$ and $S_{12}$) transmit in the first time slot, two (denoted by $S_{21}$ and $S_{22}$) transmit in the second time slot, and two (denoted by $S_{31}$ and $S_{32}$) transmit in the third time slot. Since only information about packets $S_{11}$ and $S_{12}$ could propagate to their destinations $D_{11}$ and $D_{12}$ at the end of the uplink phase, if there is only one downlink time slot, R will have no choice but to transmit $S_{21} \oplus S_{22} \oplus S_{31} \oplus S_{32}$ or $S_{21} \oplus S_{22} \oplus S_{31} \oplus S_{32} \oplus$ *some other packets* so that the other destinations could obtain information about their desired packets. We make use an observation below:

*Main Observation:* In the second time slot, we have to select two sources $S_{21}$ and $S_{22}$ whose destinations $D_{21}$ and $D_{22}$ can overhear both $S_{31}$ and $S_{32}$ in the third time slot to transmit. This is because $D_{21}$ and $D_{22}$ must be able to eliminate $S_{31} \oplus S_{32}$ from the transmission by R to obtain their desired packets $S_{21}$ and $S_{22}$, respectively.

Now, not including R, there are three possible relative position relationships between $S_{31}$ and $S_{32}$ in atom IX: 1) $S_{31}$ and $S_{32}$ are separated by three hops; 2) $S_{31}$ and $S_{32}$ are separated by two hops; and 3) $S_{31}$ and $S_{32}$ are separated by one hop. In the following, for each of the three cases, we consider whether we could find the two sources to transmit in the second time slot to satisfy the requirement as per our main observation above:

1) Without loss of generality, suppose that $S_{31}$ and $S_{32}$ are sources A and D. The destinations $D_{31}$ and $D_{32}$ are then respectively D and A. Consider the other four destinations, B, C, E, and F (their respective sources are E, F, B, and C). Destinations B and C cannot overhear D, and destination E and F cannot overhear A. None of them can overhear both sources A and D. Thus, no matter which two sources among E, F, B, and C we choose to transmit in the second time slot, we cannot satisfy the requirement of our main observation.

2) Without loss of generality, suppose that $S_{31}$ and $S_{32}$ are sources A and F. The destinations $D_{31}$ and $D_{32}$ are respectively D and C. Consider the other four destinations, A, F, E, and B (their respective sources are D, C, B, and E). Destination A cannot overhear F, destination F cannot overhear A, and destination E cannot overhear both A and F. Only destination B can overhear both A and F. Thus, no matter which two sources among D, C, B, and E we choose to transmit in the second time slot, we cannot satisfy the requirement of our main observation.

3) Without loss of generality, suppose that $S_{31}$ and $S_{32}$ are sources A and B. The destinations $D_{31}$ and $D_{32}$ are respectively D and E. Consider the four destinations, C, F, A, and B (their respective sources are F, C, D, and E). Destination C cannot overhear B, destination F cannot overhear A. Although destination A is within the transmission range of source B, they transmit in the same time slot. Since we assume half duplexity (Assumption 3), destination A cannot overhear B, and destination B cannot overhear A. None of them can overhear both sources A and B. Thus, no matter which two sources among F, C, D, and E we choose to transmit in the second time slot, we cannot satisfy the requirement of our main observation. **Q.E.D.**

**Remark for Case 1):** Even if we allow one peripheral node to transmit together with the relay in the downlink phase (at most two nodes transmit simultaneously), no matter which peripheral node transmits, at most only one of destinations among C, F, A, and B can overhear the information of both *A* and *B*. In particular, if node A (B) transmits packet A (*B*), only node B (A) overhears it and get the information of both *A* and *B*. Hence, it is still impossible to find two such destinations to satisfy the requirement of our main observation.

**Remark for Case 2):** If we allow one peripheral node to transmit together with the relay in the downlink phase (at most two nodes transmit simultaneously), the only possible way to satisfy the requirement of our main observation is to ask peripheral node B to transmit $A \oplus F$ in the downlink phase. Then destinations A and F can get both *A* and *F*. This means that in the second time slot, nodes C and D transmit; in the first time slot, nodes B and E transmit. However, by such transmission pattern, the information of packet *B* cannot propagate to its destination E in the three uplink time slots. Since node E cannot overhear both A and F, which transmit in the third timeslot, supposition 3) cannot be satisfied. Specially, if there is only one downlink time slot, R will have no choice but to transmit $A \oplus F \oplus B$ or $A \oplus F \oplus B \oplus some\ other\ packets$. It is obvious that node E cannot eliminate $A \oplus F$.

**Remark for Case 3):** Even if we allow one peripheral node to transmit together with the relay in the downlink phase (at most two nodes transmit simultaneously), only one destination among B, C, E, and F can additionally overhear the information of *D* or *A* (e.g., node B can forward the information of *A* to node F). It is still impossible to find two such destinations to satisfy the requirement of our main observation.

## APPENDIX B. INSENSITIVITY TO INNER RADIUS

The insensitivity of our simulation results to the inner radius can be explained as follows: First, when $r_i$ is small, since the nodes are randomly distributed, only a few nodes would be mapped into the area near the circle center. Second, perhaps more importantly, our problem definition requires that the distance between two peripheral nodes with traffic between them to be larger than one unit. This is to ascertain that the two nodes do indeed need the assistance of the relay R to forward the traffic between them. Consider a node A that is placed close to R. Let us draw a circle of radius one around A and another circle of radius one around R. Then the area of the latter circle that is not overlapping with the area of the former circle is the area where a node with traffic to or from A might be placed. Since A is near R, the non-overlapping area is very small. Thus a node such as A that is near R will have little traffic to be relayed by R. Hence, even if we allowed the inner radius to be small, most traffic flows would be between nodes at a distance further away from R. This explains the insensitivity of our results to the setting of the inner radius.